\documentclass[
    twocolumn,
	prl,
	amssymb,
	preprintnumbers, superscriptaddress,
	nofootinbib]{revtex4-1}

\usepackage{graphicx}
\usepackage{enumitem}
\usepackage{latexsym}
\usepackage{amsfonts}
\usepackage{amssymb}
\usepackage{xcolor}
\usepackage[export]{adjustbox}
\usepackage{amsmath}
\usepackage{slashed}
\usepackage{dcolumn}
\usepackage{verbatim}
\usepackage{float}
\usepackage{multirow}
\usepackage{xspace}
\usepackage[normalem]{ulem}
\usepackage[
pdfauthor={Djuna Croon}]{hyperref}

\newcommand{\acro}[1]{\textsc{\MakeLowercase{#1}}} 

\newcommand{\beq}{\begin{equation}}
\newcommand{\eeq}{\end{equation}}
\newcommand{\bea}{\begin{eqnarray}}
\newcommand{\eea}{\end{eqnarray}}

     \newcommand{\cL}{{\cal L}}      
 
\newcommand{\ie}{{\it i.e.}}  \newcommand{\eg}{{\it e.g.}}

\def\aj{\ref@jnl{AJ}}                   
\def\actaa{\ref@jnl{Acta Astron.}}      
\def\araa{\ref@jnl{ARA\&A}}             
\def\apj{\ref@jnl{ApJ}}                 
\def\apjl{\ref@jnl{ApJ}}                
\def\apjs{\ref@jnl{ApJS}}               
\def\ao{\ref@jnl{Appl.~Opt.}}           
\def\apss{\ref@jnl{Ap\&SS}}             
\def\aap{\ref@jnl{A\&A}}                
\def\aapr{\ref@jnl{A\&A~Rev.}}          
\def\aaps{\ref@jnl{A\&AS}}              
\def\azh{\ref@jnl{AZh}}                 
\def\baas{\ref@jnl{BAAS}}               
\def\bac{\ref@jnl{Bull. astr. Inst. Czechosl.}}
\def\caa{\ref@jnl{Chinese Astron. Astrophys.}}
\def\cjaa{\ref@jnl{Chinese J. Astron. Astrophys.}}
\def\icarus{\ref@jnl{Icarus}}           
\def\jcap{\ref@jnl{J. Cosmology Astropart. Phys.}}
\def\jrasc{\ref@jnl{JRASC}}             
\def\memras{\ref@jnl{MmRAS}}            
\def\mnras{\ref@jnl{MNRAS}}             
\def\na{\ref@jnl{New A}}                
\def\nar{\ref@jnl{New A Rev.}}          
\def\pra{\ref@jnl{Phys.~Rev.~A}}        
\def\prb{\ref@jnl{Phys.~Rev.~B}}        
\def\prc{\ref@jnl{Phys.~Rev.~C}}        
\def\prd{\ref@jnl{Phys.~Rev.~D}}        
\def\pre{\ref@jnl{Phys.~Rev.~E}}        
\def\prl{\ref@jnl{Phys.~Rev.~Lett.}}    
\def\pasa{\ref@jnl{PASA}}               
\def\pasp{\ref@jnl{PASP}}               
\def\pasj{\ref@jnl{PASJ}}               
\def\rmxaa{\ref@jnl{Rev. Mexicana Astron. Astrofis.}}%
\def\qjras{\ref@jnl{QJRAS}}             
\def\skytel{\ref@jnl{S\&T}}             
\def\solphys{\ref@jnl{Sol.~Phys.}}      
\def\sovast{\ref@jnl{Soviet~Ast.}}      
\def\ssr{\ref@jnl{Space~Sci.~Rev.}}     
\def\zap{\ref@jnl{ZAp}}                 
\def\nat{\ref@jnl{Nature}}              
\def\iaucirc{\ref@jnl{IAU~Circ.}}       
\def\aplett{\ref@jnl{Astrophys.~Lett.}} 
\def\apspr{\ref@jnl{Astrophys.~Space~Phys.~Res.}}
\def\bain{\ref@jnl{Bull.~Astron.~Inst.~Netherlands}} 
\def\fcp{\ref@jnl{Fund.~Cosmic~Phys.}}  
\def\gca{\ref@jnl{Geochim.~Cosmochim.~Acta}}   
\def\grl{\ref@jnl{Geophys.~Res.~Lett.}} 
\def\jcp{\ref@jnl{J.~Chem.~Phys.}}      
\def\jgr{\ref@jnl{J.~Geophys.~Res.}}    
\def\jqsrt{\ref@jnl{J.~Quant.~Spec.~Radiat.~Transf.}}
\def\memsai{\ref@jnl{Mem.~Soc.~Astron.~Italiana}}
\def\nphysa{\ref@jnl{Nucl.~Phys.~A}}   
\def\physrep{\ref@jnl{Phys.~Rep.}}   
\def\physscr{\ref@jnl{Phys.~Scr}}   
\def\planss{\ref@jnl{Planet.~Space~Sci.}}   
\def\procspie{\ref@jnl{Proc.~SPIE}}   

\newcommand{\samskip}{{~\\ \noindent}}

\newcommand{\Eq}[1]{Eq.~(\ref{#1})}

\allowdisplaybreaks

\begin{document}

\title{Missing in Axion: where are XENON1T’s big black holes?}
\author{Djuna Croon} \email{dcroon@triumf.ca}
\affiliation{TRIUMF, 4004 Wesbrook Mall, Vancouver, BC V6T 2A3, Canada}
\author{Samuel D.~McDermott}
\email{sammcd00@fnal.gov}
\affiliation{Fermi National Accelerator Laboratory, Batavia, IL USA}
\author{Jeremy Sakstein} \email{sakstein@hawaii.edu}
\affiliation{Department of Physics \& Astronomy, University of Hawai'i, Watanabe Hall, 2505 Correa Road, Honolulu, HI, 96822, USA}

\preprint{FERMILAB-PUB-20-270-T}

\date{\today}

\begin{abstract}
We pioneer the black hole mass gap as a powerful new tool for constraining new particles. A new particle that couples to the Standard Model---such as an axion---acts as an additional source of loss in the cores of population-III stars, suppressing mass lost due to winds and quenching the pair-instability. 
This results in heavier astrophysical black holes. As an example, using stellar simulations we show that the solar axion explanation of the recent XENON1T excess implies astrophysical black holes of $\sim 56 {\rm M}_\odot$, squarely within the black hole mass gap predicted by the Standard Model.
%
\end{abstract}

\maketitle

\noindent {\bf Introduction --} The detection of gravitational waves by the LIGO/Virgo interferometers affords us the unprecedented opportunity to examine celestial objects under a new microscope, and to test our theories of nature using the most extreme objects in the universe: black holes and neutron stars. 
The twelve mergers observed in the first two observation runs have already allowed us to verify some predictions of stellar structure theory \cite{Monitor:2017mdv}, confirm consistency with general relativity \cite{TheLIGOScientific:2016src,Sakstein:2017xjx,Baker:2017hug,Ezquiaga:2017ekz,Creminelli:2017sry}, as well as probe dark matter (\eg~\cite{Bertone:2019irm,Croon:2017zcu}).
As gravitational wave astronomy enters the realm of precision science the large volume of detected mergers will become a sensitive probe of the astrophysical populations of stellar remnants. 

In this letter, we show that black hole population studies using LIGO/Virgo can be used to probe new physics, exemplifying this by focusing on couplings between axions and electrons, as recently proposed by the XENON1T collaboration to explain the observed excess in the keV-range of electronic recoil events
\cite{Aprile:2020tmw}; in a forthcoming publication, we will study several other emission channels in models of new physics \cite{AwesomeForthcomingwork}. The XENON1T collaboration reports a $3.5\sigma$-significance for the solar axion explanation of this excess, with $g_{ae}=\mathcal{O}(10^{-12})$.
We show that such couplings affect 
the location of the black hole mass gap (BHMG) predicted by stellar structure theory. 

The BHMG is a range $45{\rm M}_\odot\le M_{\rm BH}\le 120 {\rm M}_\odot$ in which no black holes are formed by the direct collapse of massive stars due to the \emph{pair-instability}; the lower edge of the BHMG is of interest in this work as it is starting to be probed by LIGO/Virgo.
The densities and temperatures in the cores of very massive stars ($\gtrsim 50 M_\odot$) are sufficient for the production of electron-positron pairs from the plasma. These reduce the photon pressure, destabilizing the star and causing it to contract. The resulting temperature increase leads to rapid thermonuclear burning of $^{16}\textrm{O}$, which releases energy comparable to the star's binding energy. 
Stars with initial mass $\sim 50 {\rm M}_\odot$ will undergo violent pulsations, causing them to shed a large fraction of their mass before ultimately relaxing to hydrostatic equilibrium and collapsing to form a black hole. This process is referred to as a pulsational-pair instability supernova (PPISN) and results in a final black hole mass far lighter than that of its progenitor. For heavier stars, the thermonuclear explosion is so violent that the entire star becomes unbound, leaving no compact remnant behind. This process is referred to as a pair-instability supernova (PISN). For supermassive stars, the PISN is quenched due to energy losses from the photodisintegration of heavy elements such that black hole remnants reappear.

The precise location of the BHMG is sensitive to the physical processes that govern the evolution of the black hole progenitors \cite{Farmer:2019jed}: old, low metallicity, massive population-III stars. Any new physics that alters these objects could therefore change the location of the BHMG, a prediction that could be verified in the coming years \cite{Belczynski:2016jno}. Indeed, the first few LIGO/Virgo detections already indicate a lower edge of around $40 {\rm M}_\odot$,  and additional observations will provide stronger evidence \cite{Fishbach:2017zga}. 

The axion-electron coupling favored by XENON1T implies that axions are produced in the cores of stars and act as an additional source of energy loss. This alters stellar structure and evolution. In this work we display results of detailed numerical simulations of population-III stars from the zero age helium branch (ZAHB) to their ultimate collapse to form black holes or pair-instability supernovae, including such losses. From these results we derive the location lower edge of the black hole mass gap as a function of the coupling. 


\begin{figure*}[t]
    \centering
   { \includegraphics[width=0.49\textwidth]{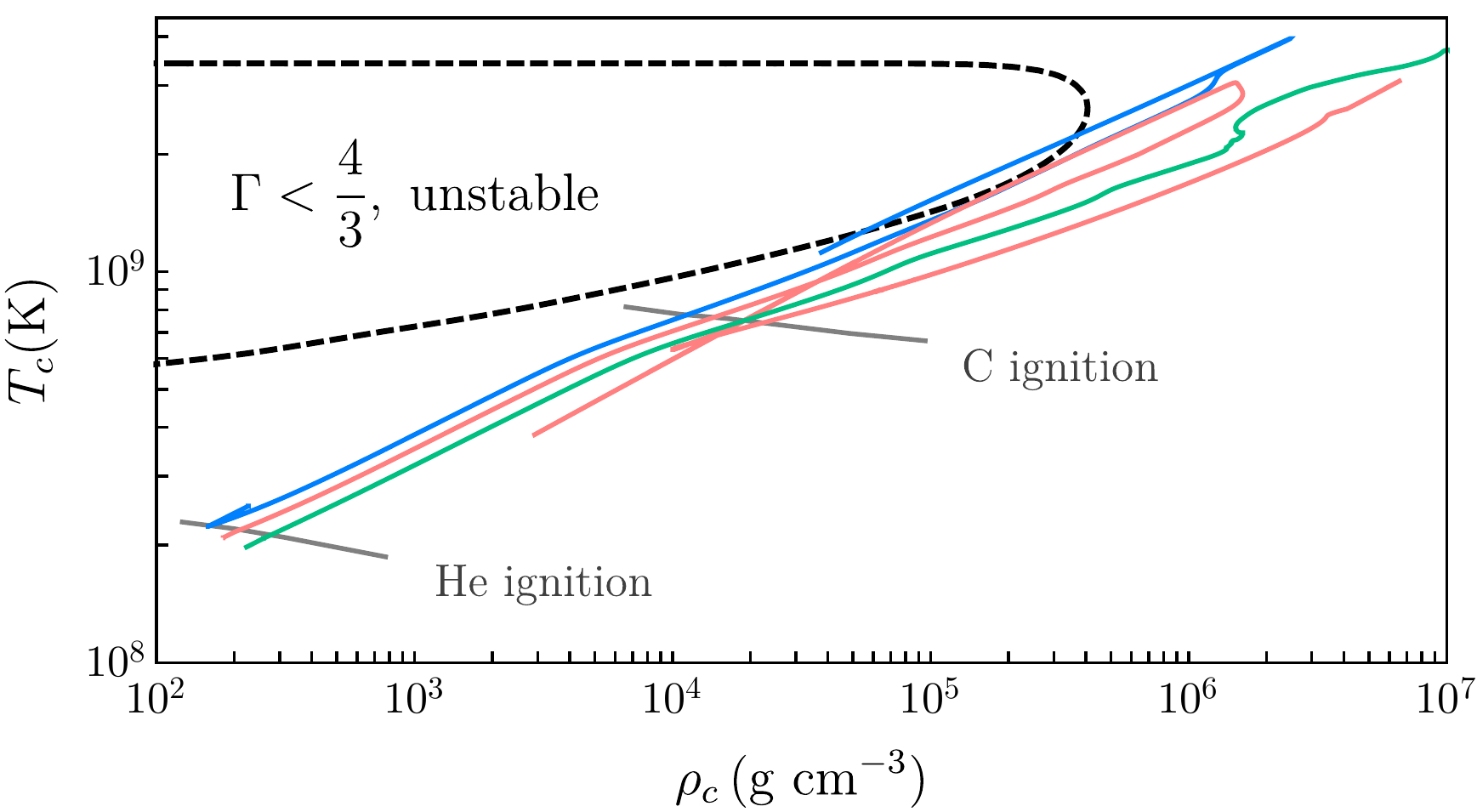}}
  { \includegraphics[width=0.49\textwidth]{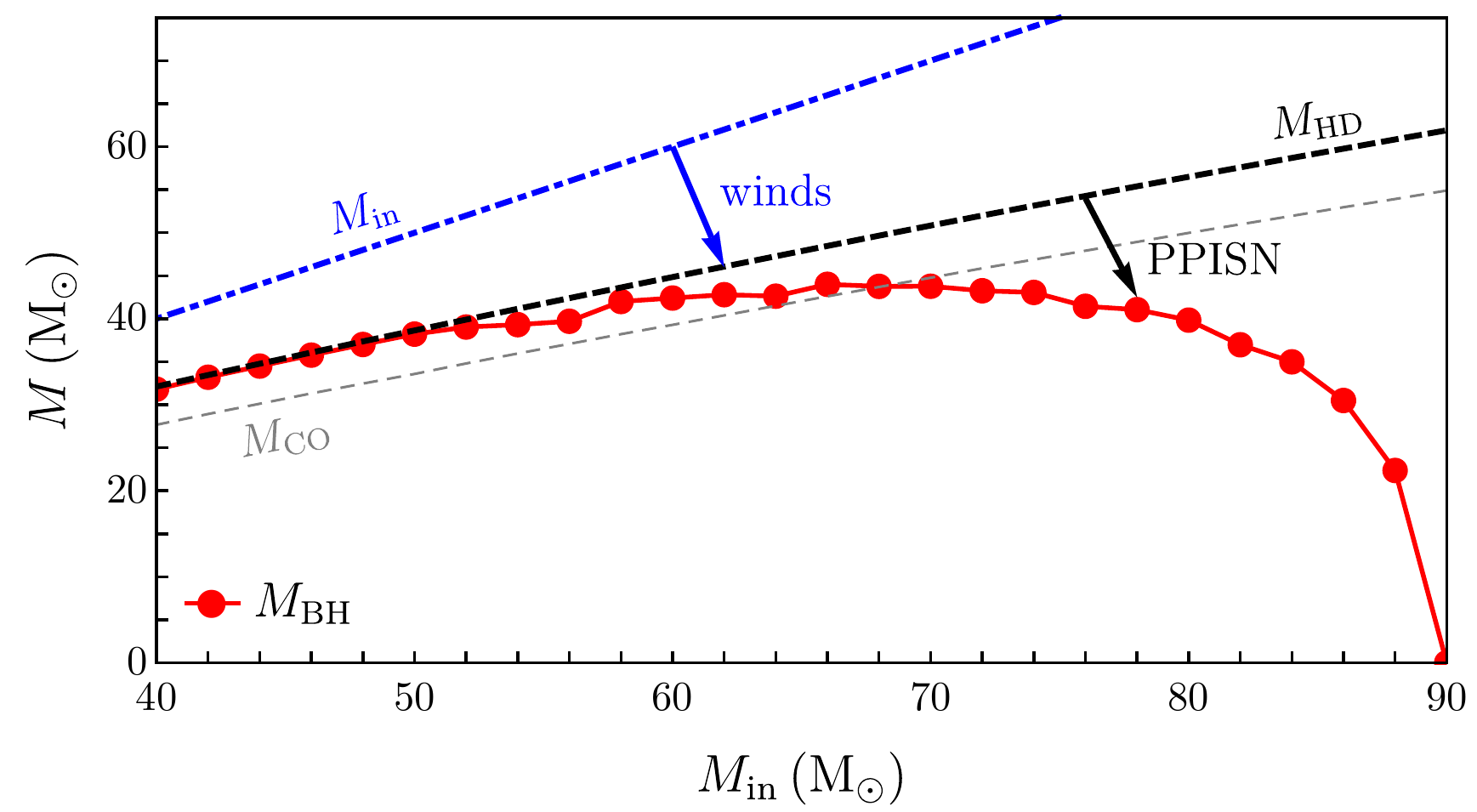}}
    \caption{
    \textbf{Left:} 
    The evolution of the central density and temperature of a 
    $M_{\rm in}=40{\rm M}_\odot$ (\textcolor[rgb]{0, 0.75, 0.5}{green}),
    $70{\rm M}_\odot$ (\textcolor[rgb]{1, 0.5, 0.5}{pink}), and $120{\rm M}_\odot$ (\textcolor[rgb]{0, 0.5, 0.99}{blue}) population-III star assuming no new physics. The region enclosed by the black dashed line indicates where the pair-instability occurs and the gray lines indicate the onset of helium and carbon burning. 
    \textbf{Right:} Various masses as a function of initial stellar mass $M_{\rm in}$ for population stars of initial metallicity ${\rm Z}_\odot/10$. The red points correspond to final black hole masses for individual stellar models. The blue dot-dashed line shows the initial mass, the black dashed line shows the mass at helium depletion, and the gray dashed line shows the CO core mass.
    }
    \label{fig:BHMG_origin}
\end{figure*}


\samskip {\bf The Pair-Instability --}
The high temperatures in the cores of population-III stars lead to the production of electron-positron pairs from the thermal plasma, through the process $\gamma\gamma\rightarrow e^+e^-$. This process has threshold energy $2m_e\simeq 10^{10}$K, but due to the long tail of the Bose-Einstein distribution, pairs will be produced when the star's temperature is $\sim 10^9$K.

The stability of stars can be expressed in terms of their equation of state (EOS). 
Stars supported by radiation pressure have EOS $\Gamma=\left(\partial P/\partial \rho\right)_s\approx 4/3.$
Stars with $\Gamma<4/3$ are unstable, so small perturbations can change their evolution drastically (see e.g. \cite{1968pss..book.....C,2012sse..book.....K}). 
We show the region (first found in~\cite{1967ApJ...148..803R}) for which pair-production causes such an instability in the $\rho$--$T$ plane in the left panel of Fig.~\ref{fig:BHMG_origin}. 
We may understand its shape by considering the rate of the pair production process. At low temperatures and densities (\ie, the lower left corner of the figure), $\gamma\gamma\rightarrow e^+e^-$ is highly Boltzmann-suppressed. As the temperature is raised, the process is less suppressed, but $e^+e^-$ pairs are produced non-relativistically.
Since the $e^+e^-$ pairs act to increase the density but not the pressure, they effectively lower the volume-averaged $\Gamma$ below $4/3$.
At temperatures $T \gtrsim m_e$, the process $\gamma \gamma \to e^+e^-$ becomes unsuppressed, and the $e^+e^-$ are relativistic; their EOS is now $\Gamma=4/3$ and they contribute significant pressure, removing the instability. The termination of the instability at high densities occurs because increasing the density increases the pressure due to ions, which follow the ideal gas law $P\propto\rho T$ and have EOS $\Gamma_{\,\textrm{ions}}=5/3$, causing them to dominate the EOS. 


\samskip {\bf Physical Origin of the BHMG --} 
The physical origin of the BHMG can be understood by considering the evolution of population-III stars from the onset of helium burning through to core collapse, illustrated in Fig.~\ref{fig:BHMG_origin}. The left panel shows the evolution of the central temperature $T_c$ and density $\rho_c$ of three stars from the ZAHB (in the bottom left of the figure) through the instability region. From the ZAHB, all three stars start to burn helium, evolving to higher densities and temperatures. During this time, they lose mass due to stellar winds. After the onset of carbon burning, the three tracks diverge. The $40{\rm M}_\odot$ star does not encounter the instability and continues on its original trajectory, eventually collapsing to form a black hole. The $70{\rm M}_\odot$ star does become unstable, evolving into a PPISN and shedding mass during pulsations. Eventually, this star relaxes to hydrostatic equilibrium and continues its evolution until it core-collapses to form a black hole. Lastly, the $120{\rm M}_\odot$ star experiences such violent pulsations that the entire star undergoes a PISN, leaving no compact remnant. 

The influence of these processes on the black hole mass distribution is shown in the right panel of Fig.~\ref{fig:BHMG_origin} where we plot the final black hole mass as a function of initial mass for stars with metallicity ${\rm Z}_\odot/10$ (where ${\rm Z}_\odot = 0.0142$). The blue dot-dashed line shows the initial mass. This would be the final mass of the black hole if it were not for the stellar winds during the helium burning phase. The mass remaining after helium depletion is shown by the black dashed line. The stellar core at this time is comprised primarily of ${}^{12}$C and ${}^{16}$O; this mass is shown in the thin gray dashed line.

Low mass stars (\eg~the $40{\rm M}_\odot$ star in the left panel) that avoid the instability form black holes whose mass is nearly identical to their mass at helium depletion (denoted $M_{\rm HD}$ in the right panel), while higher mass stars (\eg~the $70{\rm M}_\odot$ star in the left panel) experience further mass loss due to the PPISN. Increasingly heavy stars experience increasingly larger fractional mass losses. At some point, the mass lost due to the PPISN is so large that the final black hole is lighter than would have been formed from a less massive star; this is visible in the turnover seen in the right panel. Eventually, the stars are heavy enough to undergo a PISN, at which point no black hole is formed. In the case of Fig.~\ref{fig:BHMG_origin}, this happens around an initial mass of $M=90{\rm M}_\odot$. 

The heaviest mass black hole that can be formed due the competition between the initial mass and the mass lost due to the PPISN gives the lower edge of the black hole mass gap. The lightest mass black hole formed after the PISN is quenched corresponds to the upper edge. The intervening region, where the prediction of Standard Model astrophysical processes is an absence of black holes, is the BHMG. Black holes formed in previous mergers may exist within the BHMG, but the rate for such mergers is necessarily smaller than those from direct stellar antecedents \cite{Mangiagli:2019sxg}.


\samskip {\bf Effects of the Axion-Electron Coupling --}
We consider a model of an ``electrophilic axion'' in which the Standard Model is supplemented with 
\begin{equation} \label{Lae}
    \cL_{ae} = - i g_{ae}\bar\psi_e \gamma_5 \psi_e a,
\end{equation}
where $g_{ae}$ is a dimensionless coupling\footnote{As axions are pseudo-Goldstone Bosons, this coupling stems from a derivative interaction suppressed by a mass scale $\Lambda$: $\partial_\mu a\, \bar\psi_e \gamma^\mu \gamma_5 \psi_e/\Lambda$, where we expect $g_{ae} \sim \mathcal O(m_e/\Lambda)$.}, $\psi_e$ is the electron Dirac field, and $a$ is an axion-like particle (henceforward axion). 
For convenience, we define $\alpha_{26} \equiv 10^{26} g^2_{ae}/4\pi$.
We assume the axion is massless, which is valid for $m_{a} \ll {\rm keV}$. For nonzero $\alpha_{26}$, light axions will be produced during different stages of the stellar evolution. The most consequential impacts are during helium burning.
The XENON1T collaboration reports an excess that can be fit with the interaction of \Eq{Lae} and a value of $\alpha_{26}\in\{55,109\}$ at 90\% CL \cite{Aprile:2020tmw}. This may appear to be in conflict with constraints on stars in other systems \cite{DiLuzio:2020jjp}, but unexplored parameter degeneracies \cite{1983A&A...128...94B} may substantially impact these bounds. Given these uncertainties, independent probes of this parameter space are warranted. Furthermore, stellar population bounds are unlikely to improve substantially in the near future, whereas BH population studies are in their infancy, with a large volume of high-precision data on the horizon.

We consider losses due to axion emission at temperatures below $10^9$K, such that electrons are non-relativistic. In that limit, the specific loss rate for semi-Compton scattering, $e+\gamma \rightarrow e+a$ is given by \cite{Raffelt:1994ry},
\begin{eqnarray}\label{eq:Q_SC}
    \mathcal Q_{\rm sC} = \frac{40 \,\zeta_6 \alpha_{\rm EM} g_{ae}^2}{\pi^2}  \, \frac{Y_e T^6}{m_N m_e^4} \, F_{\rm deg} \simeq
    33\alpha_{26}Y_eT_8^6F_{\rm deg} {\rm \frac{erg}{g \!\cdot\! s} },~~~~
\end{eqnarray}
where $\alpha_{\rm EM} = 1/137$ is the electromagnetic fine-structure constant, $m_{N,e}$ are the nucleon and electron mass respectively, 
$Y_e = (Z/A) $ is the number of electrons per baryon, $T_8=(T/10^8\textrm{K})$, and $\zeta_6 = \pi^6/945$. The function $F_{\rm deg}$ encodes the Pauli-blocking of the process due to electron degeneracy: $
    F_{\rm deg} = 2 n_e^{-1} \int d^3 \mathbf{p}\,(2 \pi)^{-3} f_{e^-} (1-f_{e^-})$, where $f_{e^-} = [e^{(E-\mu)/T}+1]^{-1}$ is the $e^-$ distribution function.
We find that a good approximation of $F_{\rm deg}$ is
\begin{align} \notag
    F_{\rm deg} &= \frac{1}{2} \left[ 1-\tanh  f(\rho,T) \right] \\ f(\rho,T) &=
    a \log_{10} \left[ \frac{\rho}{\text{g cm}^{-3}}\right] -b \log_{10}\left[ \frac{T}{\rm K}\right] +c,
\end{align}
with coefficients $a=0.973$, $b=1.596$, and $c=8.095$. During stellar helium burning, we find $F_{\rm deg} \approx 1$.

The specific energy loss due to bremsstrahlung $e+ (Z,A) 
\to e + (Z,A) + a$ depends on the nucleon degeneracy. For nonrelativistic electrons, the rate in the non-degenerate (ND) and degenerate (D) regimes is \cite{Raffelt:1994ry}
\begin{eqnarray}
    \label{eq:Q_bremm}
    &&\mathcal Q_{b,{\rm ND}} \!=\! \frac{128}{45} \frac{\alpha_{\rm EM}^2 \alpha_{26} \rho T^{5/2}}{\sqrt{\frac\pi2} m_N^2 m_e^{7/2}} F_{b,{\rm ND}}  \simeq 582 \, \alpha_{26} {\rm \frac{erg}{g \! \cdot \! s} } \rho_6 T_8^{5/2} F_{b,{\rm ND}} \nonumber
    \\ 
    &&\mathcal Q_{b,{\rm D}} \!=\!\frac{\pi^2}{15}\frac{Z^2}{A}\frac{\alpha_{\rm EM}^2 \alpha_{26} T^4}{m_N m_e^2} F_{b,{\rm D}} \simeq 10.8\, \alpha_{26} {\rm \frac{erg}{g \! \cdot \! s} }  T_8^4 F_{b,{\rm D}}
\end{eqnarray}
where $\rho_6 = \rho/(10^6{\rm g/cm^3})$, $F_{b,{\rm ND}}  = Z(1+Z)/A$ 
for metallicity $Z$,
and
$F_{b,{\rm D}} = \frac{2}{3} \log\left(1+2\kappa^{-2} \right) + \left[ (\kappa^2 +2/5) \log\left( 1+2\kappa^{-2} \right) - 2\right] \beta_F^2/3$ to second order in the velocity at the Fermi surface $\beta_F = p_F/E_F$,
with the Debye scale $\kappa^2 = 2\pi \alpha_{\rm EM} n_e/(Tp_F^2).$
The total specific loss rate is $\mathcal Q = \mathcal Q_{\rm sC} + ( \mathcal Q_{b,{\rm ND}}^{-1} + \mathcal Q_{b,{\rm D}}^{-1})^{-1}$  \cite{Raffelt:1994ry}.
The effects of axio-recombination and -deexcitation are negligible in population-III stars because these processes are suppressed in low-metallicity objects \cite{Redondo:2013wwa}. The semi-Compton rate dominates at low density, so we expect this to determine the axionic energy loss rate in the early phases of the stellar evolution.

These axion energy emission rates directly impact the BHMG by speeding the evolution of massive stars. 
As explored in more detail in \cite{AwesomeForthcomingwork}, this follows because increased energy loss hastens the rate of helium depletion. The effect of this is twofold. First, there is less time to lose mass to stellar winds, and, second, the amount of carbon fused to oxygen during the helium burning phase is reduced. This alters the dynamics and interplay of C/O burning, ultimately reducing the violence of pulsations, leading to a heavier black hole mass \cite{Farmer:2020xne}.


\begin{figure*}
    \centering
    {\includegraphics[width=1.\textwidth]{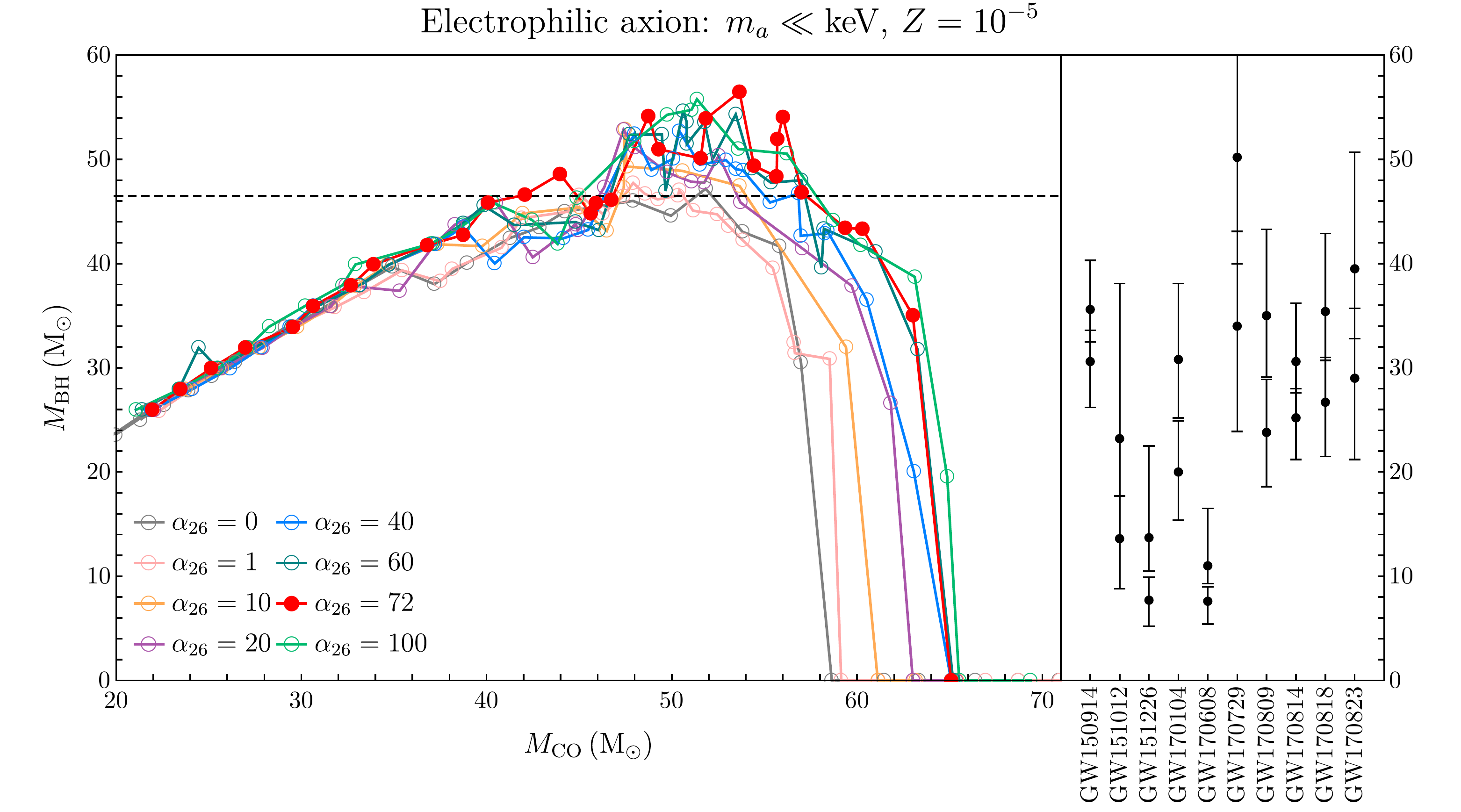}}
    \caption{Final black hole mass as a function of CO 
    core mass for different values of $\alpha_{26}$. The emphasized value $\alpha_{26}=72$ 
    provides a good fit to the XENON1T excess \cite{Aprile:2020tmw} and produces the largest mass black hole in our simulations.}
    \label{fig:grid_axions}
\end{figure*}

\samskip {\bf Stellar Modeling --}
We simulate the evolution of the black hole progenitors using the stellar structure code {\tt{MESA}} version 12778 (\cite{Paxton:2010ji,*Paxton:2013pj,*Paxton:2015jva,*Paxton:2017eie}) updated to include the losses due to axion emission given in equations~\eqref{eq:Q_SC}--\eqref{eq:Q_bremm}. Our prescription for simulating the PPISN, PISN, and core collapse collapse follows that of references \cite{Marchant:2018kun} and \cite{Farmer:2019jed}. Specifically, we use the wind prescription of \cite{Brott:2011ni}: $\dot{M}\propto(Z/{\rm Z}_\odot)^{0.85}$ with ${\rm Z}_\odot = 0.0142$. Convection is modelled using mixing length theory \cite{1968pss..book.....C} with efficiency parameter $\alpha_{\rm MLT}=2.0$ and semi-convection is modelled using the prescription of \cite{1985A&A...145..179L} with efficiency $\alpha_{\rm SC}=1.0$. Connvective overshooting is exponential with $f_0=0.005$ and $f_{\rm ov}=0.01$ (see \cite{Farmer:2019jed} for the definition of these parameters).

Our code follows the star's evolution from the ZAHB to either core collapse or PISN. We begin by forming an initial helium star of mass $M$, metallicity $Z$, and helium-4 fraction $Y({^4{\rm He}})=1-Z$. Following \cite{Marchant:2018kun,Farmer:2019jed} we define helium depletion as the point where the central helium mass fraction falls below $0.01$, and the mass of the carbon-oxygen core, $M_{\rm CO}$, as the mass interior to the point where the helium mass fraction is larger than $0.01$. We define the mass of the black hole as the mass of material at core collapse with velocities smaller than the escape velocity $v_{\rm esc}(r)=\sqrt{GM(r)/r}$.

\begin{figure}
    {\includegraphics[width=.49\textwidth]{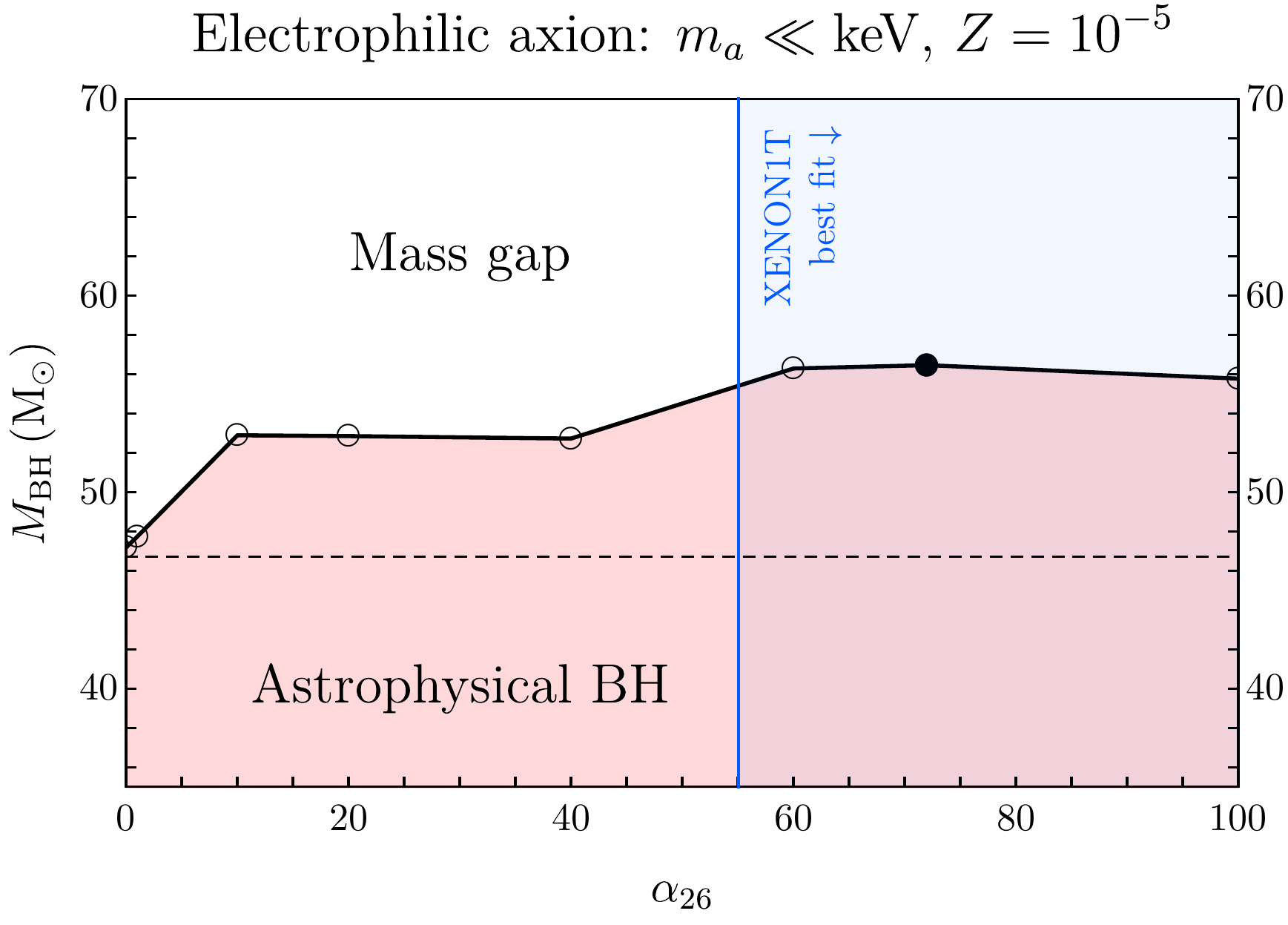}}
    \caption{Lower edge of the BHMG for different values of $\alpha_{26}$. The dashed line corresponds to the lower edge of the mass gap for  $\alpha_{26}=0$, consistent with \cite{Farmer:2019jed}. The blue shaded region denotes the XENON1T fit \cite{Aprile:2020tmw}.}
    \label{fig:massgap_axions}
\end{figure}

\samskip {\bf The BHMG and leptophilic axions --}
We have computed grids of stars with initial masses between $20{\rm M}_\odot$ and $90{\rm M}_\odot$ in steps of $1{\rm M}_\odot$. We take $Z=10^{-5}$, which is representative of population-III stellar progenitors. This also leads to less conventional mass loss, and therefore gives the lower edge of the BHMG found in \cite{Farmer:2019jed}
\footnote{As reviewed in \cite{Farmer:2019jed}, Standard Model astrophysics provides several sources of uncertainty on the precise boundary of the BHMG. Most of the uncertainties marginalized over changed the standard value $M_{\rm BH} = 45 {\rm M}_\odot$ by less than $4 {\rm M}_\odot$, with the exception of the reaction ${}^{12}$C$(\alpha,\gamma)^{16}$O \cite{Farmer:2020xne}, which we will explore in \cite{AwesomeForthcomingwork}. 
}.

In the left panel of Fig.~\ref{fig:grid_axions} we show the resulting BH remnant mass as a function of the carbon-oxygen (CO) core mass.
We stress that the CO core mass is not observable, and used for visualization purposes only. 
We reproduce the result $M_{\rm BH} = 46 {\rm M}_\odot$ of \cite{Farmer:2019jed} in the $\alpha_{26} = 0$ limit, and no significant deviation from this value is observed for $\alpha_{26} = 1$. The situation is different for higher values of $\alpha_{26}$. An excursion to a higher mass black hole is visible for $\alpha_{26} = 10$, and final black hole masses above $56{\rm M}_\odot$ are possible for $\alpha_{26} \gtrsim 50$. We summarize the results of the left panel of Fig.~\ref{fig:grid_axions} in Fig.~\ref{fig:massgap_axions}, where we show the maximum value of $M_{\rm BH}$ attained over the grid of initial masses as a function of $\alpha_{26}$. One can observe that electrophilic axions in the XENON1T preferred parameter space can produce large black holes deep in the BHMG.

In the right panel of Fig.~\ref{fig:grid_axions} we show black hole masses derived from waveform analysis in binary merger events.
Interestingly, one event, GW170829, involves a black hole with mass $50.2_{10.2}^{16.2}{\rm M}_\odot$. The large error bars make this consistent with both XENON1T's best fitting axion model and the absence of axions. Reference \cite{Fishbach:2017zga} have used 10 events to find evidence that the lower edge of the gap lies around $40{\rm M}_\odot$, in strong tension with the predictions of electrophilic axions. In a forthcoming publication \cite{AwesomeForthcomingwork}, we intend to perform a similar analysis incorporating losses from other types of novel particles, using the upcoming LIGO/Virgo O3 data release, and considering new observational signatures and detection strategies.


\samskip {\bf Discussion and Outlook --} 
The advent of gravitational wave astronomy has opened a new window into the inner workings of the Universe. The handful of gravitational wave events we have observed to date have already expanded our knowledge of the laws governing the physical processes that play out on the cosmic stage. As LIGO/Virgo's third observing run is concluded, the apparatus is upgraded to even higher sensitivities, and future detectors come online, we expect tens to thousands of black hole merger events per year, bringing gravitational wave astronomy into the realm of precision science. It is of paramount importance for maximizing the discovery potential of the data that we identify the physical observables that can discriminate between competing theories of fundamental physics.

In this work we have presented one such novel probe: the location of the lower edge of the black hole mass gap. We have demonstrated that losses due to new light particles increase the mass of astrophysical black hole remnants by shortening the duration of helium burning, which reduces the mass lost due to stellar winds and lowers the amount of combustible oxygen, which suppresses the pulsational pair instabilities experienced by their population-III 
progenitors. We exemplified this using the electrophilic axion, finding that black holes with masses as large as $56{\rm M}_\odot$ can be formed. These would lie deep inside the mass gap predicted by the standard model, whose lower edge lies at $\sim 46{\rm M}_\odot$. Indeed, one interpretation of the recent excess observed by XENON1T \cite{Aprile:2020tmw} is that it is due such particles produced in the sun. If this excess persists then the observation of a population of heavy black holes can confirm this hypothesis.



\samskip {\it Note Added:} While this work was in preparation, related works on the axion interpretation of the XENON1T excess \cite{DiLuzio:2020jjp,
Sun:2020iim,*Cacciapaglia:2020kbf,*Dessert:2020vxy,*Coloma:2020voz,*Dent:2020jhf,*Bloch:2020uzh,*Gao:2020wer,*Budnik:2020nwz,*OHare:2020wum,*Takahashi:2020bpq} appeared, as did
\cite{Baek:2020owl,*Hryczuk:2020jhi,*Ko:2020gdg,*Gao:2020wfr,*Chao:2020yro,*Ge:2020jfn,*Bhattacherjee:2020qmv,*DelleRose:2020pbh,*An:2020tcg,*McKeen:2020vpf,*DeRocco:2020xdt,*Chala:2020pbn,*Lindner:2020kko,*Zu:2020idx,*An:2020bxd,*Baryakhtar:2020rwy,*Bramante:2020zos,*Jho:2020sku,*Gelmini:2020xir,*Nakayama:2020ikz,*Primulando:2020rdk,*Khan:2020vaf,*Robinson:2020gfu,*Cao:2020bwd,*Lee:2020wmh,*Paz:2020pbc,*AristizabalSierra:2020edu,*Choi:2020udy,*Buch:2020mrg,*Bell:2020bes,*Chen:2020gcl,*Dey:2020sai,*Du:2020ybt,*Su:2020zny,*Bally:2020yid,*Harigaya:2020ckz,*Boehm:2020ltd,*Amaral:2020tga,*Fornal:2020npv,*Alonso-Alvarez:2020cdv} focusing on other possible explanations.

\samskip {\bf Acknowledgments --}
Many thanks to Eric Baxter, Kristina Launey, Jess McIver, Jason Kumar, David McKeen, Samaya Nissanke, Noemi Rocco, and Istvan Szapudi for useful conversations. We are grateful to the MESA community for answering our many questions.
T\acro{RIUMF} receives federal funding via a contribution agreement with the National Research Council Canada.
Fermilab is operated by Fermi Research Alliance, LLC under Contract No. De-AC02-07CH11359 with the United States Department of Energy.

\appendix

\bibliography{refs}

\end{document}